\documentclass[
  journal=medium,
  manuscript=article-type,
  year=2020,
  volume=37,
]{cup-journal}

\usepackage{amsmath}
\usepackage[nopatch]{microtype}
\usepackage{booktabs}
\usepackage{graphicx}
\usepackage{pgf}
\usepackage{color}
\usepackage{amssymb}
\usepackage{amsfonts}
\usepackage[algo2e]{algorithm2e} 
\usepackage{algorithm}
\usepackage{verbatim}
\usepackage{caption}
\usepackage{subcaption}
\usepackage{url}
\usepackage{setspace}

\newcommand{\bc}{\boldsymbol c}

\newcommand{\bA}{{\bf A}}
\newcommand{\bP}{{\bf P}}
\newcommand{\btheta}{\boldsymbol\theta}

\title{A label-switching algorithm for fast core-periphery identification}

\author{Eric Yanchenko}
\affiliation{Human and AI Center, Akita International University, Akita, Japan}
\email[Eric Yanchenko]{eyanchenko@aiu.ac.jp}

\author{Srijan Sengupta}
\affiliation{Department of Statistics, North Carolina State University, Raleigh, NC, USA}
 %% First letter not capped

\begin{document}

\begin{abstract}
Core-periphery (CP) structure is frequently observed in networks where the nodes form two distinct groups: a small, densely interconnected core and a sparse periphery. Borgatti and Everett (2000) proposed one of the most popular methods to identify and quantify CP structure by comparing the observed network with an ``ideal'' CP structure. While this metric has been widely used, an improved algorithm is still needed. In this work, we detail a greedy, label-switching algorithm to identify CP structure that is both fast and accurate. By leveraging a mathematical reformulation of the CP metric, our proposed heuristic offers an order-of-magnitude improvement on the number of operations compared to a naive implementation. We prove that the algorithm monotonically ascends to a local maximum while consistently yielding solutions within 90\% of the global optimum on small toy networks. On synthetic networks, our algorithm exhibits superior classification accuracies and run-times compared to a popular competing method, and on one-real world network, it is 340 times faster.
\end{abstract}

\section{Introduction}
Networks are a simple model for complex relationships with entities or objects represented as {\it nodes} and their interactions or relationships encoded as {\it edges}. One observed feature in many networks is {\it core-periphery} (CP) structure, where nodes form two distinct groups: a densely-connected core and a sparsely-connected periphery \citep{Csermely2013, yanchenko2022coreperiphery}. While there is not a universal definition, CP structure is distinct from the more well-known assortative mixing in that the core and periphery are densely and sparsely connected, respectively, and there is greater density between core and peripheral nodes than within the periphery itself. CP structure has been observed across many domains including: airport networks \citep{lordan2017analyzing, lordan2019core}, power grid networks \citep{yang2021optimizing}, economic networks \citep{krugman1996self, magone2016core}, and more. We refer the interested reader to \citep{yanchenko2022coreperiphery} for a more thorough review.

One of the most popular approaches to detect and quantify CP structure is the Borgatti and Everett (BE) metric which finds the correlation between the observed network and an ``ideal'' CP structure \citep{BORGATTI2000}. Here, we focus on the discrete model where nodes are assigned exclusively to either the core or periphery. There are multiple implementations available for this method, including the UCInet software \citep{everett2002ucinet} and \texttt{cpnet} Python package. But as networks continue to increase in size, it is important for applied researchers to have efficient algorithms to quickly and accurately detect this CP structure. 

In this work, we detail a greedy, label-switching algorithm to identify CP structure in networks. Our first contribution is to thoroughly benchmark an existing algorithm \citep{yanchenko2022divide}, demonstrating that it yields more accurate CP labels compared to a competing implementation. Moreover, we prove that this algorithm enjoys both monotone ascent and finite termination at a local optimum, and, on small networks, typically yields objection function values within 90\% of the global optimum. The other major contribution is a computational improvement to the algorithm in \cite{yanchenko2022divide}, introducing an incremental-update implementation which significantly improves speed and scalability. Specifically, we exploit a mathematical reformulation of the BE metric which yields an order-of-magnitude speed-up compared to a naive algorithm. Empirical results show the algorithm to be significantly faster than both a naive implementation and competing method, and on one real-world network (DBLP \citep{gao2009graph, ji2010graph}), it is 340 times faster.

In related work, there have been several notable developments in fast algorithms for statistical network analysis, particularly for community detection, model fitting, and two-sample testing. \cite{amini2013} proposed a fast pseudo-likelihood method for community detection, which was subsequently refined by \cite{wang2021fast}. \cite{zhang2022distributed} introduced a distributed algorithm for large-scale block models with a grouped community structure. Divide-and-conquer strategies have also gained attention as scalable alternatives to global community detection in large networks \citep{mukherjee2021two, wu2020distributed, chakrabarty2023sonnet}. More recently, subsampling-based approaches have been developed for computationally efficient network analysis. \cite{bhadra2025scalable} proposed a predictive subsampling method for fast community detection, while \cite{chakraborty2025scalable} introduced a subsampling-based framework for scalable estimation and two-sample testing. In parallel, \cite{chakrabarty2025network} developed a subsampling-based approach for network cross-validation and model selection. Lastly, \cite{yanchenko2022divide} and \cite{yanchenko2025graph} develop divide-and-conquer algorithms for fast CP identification in large networks. Here, we leverage the base algorithm of these works and further detail the computational improvements.

The remainder of this paper is organized as follows. In Section \ref{sec:method}, we describe the BE metric while in Section \ref{sec:alg} we detail a computationally efficient algorithm and examine some of its properties.  Experiments on simulated and real-world networks are the subjects of Sections \ref{sec:sim} and \ref{sec:real}, respectively, and we share concluding thoughts in Section \ref{sec:conc}.

\section{Core-periphery metric}\label{sec:method}
Consider a network with $n$ nodes and let $\bA$ be the corresponding $n\times n$ adjacency matrix where $A_{ij}=1$ if nodes $i$ and $j$ have an edge, and 0 otherwise, and $A_{ij}\mid P_{ij}\stackrel{\text{ind.}}{\sim}\mathsf{Bernoulli}(P_{ij})$ for $0\leq i<j\leq n$. We write $\bA\sim\bP$ as shorthand for this model. For simplicity, we only consider unweighted, undirected networks with no self-loops. We also define $\bc\in\{0,1\}^n$ as the CP assignment vector where $c_i=1$ if node $i$ is assigned to the core, and 0 if assigned to the periphery where $k=\sum_{i=1}^n c_i$ is the size of the core.
For a given $\bc$, let $\Delta(\bc)\equiv \Delta$ be an $n\times n$ matrix such that $\Delta_{ij}=c_i+c_j - c_ic_j$, i.e., $\Delta_{ij}=1$ if node $i$ or $j$ is assigned to the core, and 0 otherwise. Then $\Delta$ represents an ``ideal'' CP structure where each core node is connected to all other nodes, each periphery node is connected to all core nodes, but periphery nodes have no connections among themselves. The notion of ideal CP structure can be generalized such that $\Delta_{ij}\in(0,1)$ or $\Delta_{ij}=N/A$ when one node is in the core and the other in the periphery \citep{BORGATTI2000}.

Although there are many possible definitions of CP structure, \cite{BORGATTI2000} quantify the property by defining:
\begin{equation}\label{eq:be}
    T(\bA,\bc)
    =\mathsf{Cor}(\bA,\Delta),
\end{equation}
i.e., the Pearson correlation between the upper-triangles of $\bA$ and $\Delta$. We can re-write \eqref{eq:be} as
\begin{align}\label{eq:metric}\notag
    T(\bA,\bc)
    &=\frac{\sum_{i<j}(A_{ij}-\bar A)(\Delta_{ij}-\bar \Delta)}{\frac12n(n-1)\{\bar A(1-\bar A)\bar\Delta(1-\bar\Delta)\}^{1/2}}\\
    &=\frac{\sum_{i<j}A_{ij}\Delta_{ij}-\frac12n(n-1)\bar A\bar\Delta}{\frac12n(n-1)\{\bar A(1-\bar A)\bar\Delta(1-\bar\Delta)\}^{1/2}}
\end{align}
where $\bar X$ is the entry-wise mean of the upper-triangle of matrix $X$. For a recent extension of this metric, please see \cite{estevez2025revising}. The form in \eqref{eq:metric} is more amenable to statistical analysis and will be leveraged in the remainder of this work. Lastly, the labels $\bc$ are typically unknown so we must find the optimal labels, i.e.,
\begin{equation}\label{eq:opt}
    \hat\bc
    =\arg\max_{\bc}\{T(\bA,\bc)\}.
\end{equation}
The space of all possible solutions, $\{0,1\}^n$, has $\mathcal O(e^n)$ elements, meaning that an exhaustive search to find the globally optimal solution is not feasible for even moderately sized networks. Thus, an optimization algorithm must be employed to find an approximation of the global optimum. The remainder of this paper is devoted to such an algorithm that is both computationally efficient and empirically accurate.

\section{Algorithm}\label{sec:alg}

\subsection{Label-switching algorithm}

We first detail an efficient algorithm to approximate $\hat\bc$ in \eqref{eq:opt}, building off that of \cite{yanchenko2022divide}. While both \cite{yanchenko2022divide} and \cite{yanchenko2025graph} employ divide-and-conquer methods, our focus is working directly with the entire network, though the following algorithm can easily be integrated into these approaches. After randomly initializing the CP labels, one at a time and for each node $i$, we swap the label of the node and compute \eqref{eq:metric} with these new labels. The new label is kept if the objective function value is (strictly) larger than before, otherwise the original label remains. This process repeats until the labels are unchanged for an entire cycle through all $n$ nodes. Our approach is similar to the Kernighan-Lin algorithm for graph partitioning \citep{Kernighan1970} since both perform greedy, local searches. Unlike our algorithm, however, the group sizes are fixed in the Kernighan-Lin algorithm. The proposed approach is detailed in Algorithm 1.

\begin{algorithm}[H]
\SetAlgoLined
\KwResult{Core-periphery labels $\bc\in\{0,1\}^n$}
 {\bf Input: } adjacency matrix $\bA\in\{0,1\}^{n\times n}$\;
 
 Initialize labels $\bc$\;
 
 $run = 1$\;
 
 \While{$run > 0$}{
 
 Set $run =0$\; 
 
 Randomly order nodes\;
 
  \For{$i$ in $1:n$}{
  $\bc'=\bc$, $c'_i=1-c_i$\;
  
  \If{$T(\bA, \bc') > T(\bA, \bc)$}{
  
  $\bc\longleftarrow\bc'$ \; 
  
  $run = 1$\;
 }
   
  }
 }
 \caption{Label-switching for core-periphery detection}
 \label{alg:greedy}
\end{algorithm}

\subsection{Computational improvement}
The main computational bottleneck in Algorithm 1 comes from computing $T(\bA,\bc')$. Ostensibly, this requires $\mathcal O(n^2)$ operations since we must compute $M(\bc)=\sum_{i<j}A_{ij}\Delta_{ij}(\bc)$ and $\bar A=\sum_{i<j}A_{ij}/\{\frac12n(n-1)\}$. We call such an implementation \texttt{greedyNaive}. We can leverage the re-formulation of the BE metric from \eqref{eq:metric}, however, to improve the computational efficiency of Algorithm 1 from that of the original algorithm in \cite{yanchenko2022divide}. The following is a similar idea to that of \cite{kojaku2017}. In particular, $T(\bA,\bc')$ can be efficiently updated since $\bc'$ differs from the current CP labels by only a single entry. While the following approach has been used in previous works \citep[e.g.,][]{yanchenko2025graph}, the details of this computational improvement have yet to be formally described and thoroughly investigated.

We call the following approach \texttt{greedyFast}. At initialization, there are no shortcuts and the objective function $T(\bA,\bc)$ must be computed as in \eqref{eq:metric}, which takes $\mathcal O(n^2)$ operations. Now, let $(\bc')^{(i)}$ be the proposed CP labels where $(c')_i^{(i)}=1-c_i$ and $(c')_j^{(i)}=c_j$ for $j\neq i$, i.e., the labels differ only at node $i$. It is clear that we only need to update $\bar\Delta$ and $M(\bc)$ to compute $T(\bA,(\bc')^{(i)})$. We know that $\bar\Delta(\bc)=\{\frac12k(k-1)+k(n-k)\}/\{\frac12n(n-1)\}$ where $k=\sum_{j=1}^n c_j$. If $k'$ is the size of the core for labels $(\bc')^{(i)}$, then $k'=k+1$ if $c_i=0$, and $k'=k-1$ if $c_i=1$. Thus $\bar\Delta((\bc')^{(i)})=\{\frac12k'(k'-1)+k'(n-k')\}/\{\frac12n(n-1)\}$ can be computed in $\mathcal O(1)$. The key step is computing $M((\bc')^{(i)})$ by using $M(\bc)$. Notice that 
\begin{align}\label{eq:comp}\notag
    M((\bc')^{(i)})
    &=\sum_{j=1}^n\sum_{k=1}^{j-1} A_{jk}\Delta_{jk}((\bc')^{(i)}) \\ \notag 
    &=\sum_{j\neq i}^n\sum_{k=1}^{j-1}A_{jk}\Delta_{ik}((\bc')^{(i)})
    +\sum_{k=1}^{i-1} A_{ik}\Delta_{ik}((\bc')^{(i)})\\ \notag
    &=\sum_{j\neq i}^n\sum_{k=1}^{j-1}A_{jk}\Delta_{jk}(\bc) 
    +\sum_{k=1}^{i-1} A_{ik}\Delta_{ik}((\bc')^{(i)})\\ \notag
    &= \left(M(\bc) -\sum_{k=1}^{i-1} A_{ik}\Delta_{ik}(\bc)\right) +\sum_{k=1}^{i-1} A_{ik}\Delta_{ik}((\bc')^{(i)})\\
    &=M(\bc)+\sum_{k=1}^{i-1}A_{ik}\{\Delta_{ik}((\bc')^{(i)})-\Delta_{ik}(\bc)\}.
\end{align}

\noindent
Since we have already computed$M(\bc)$, the only term we need to compute in \eqref{eq:comp} is the final sum, which requires $\mathcal O(n)$ operations. Thus, while the initial evaluation of $T(\bA,\bc)$ is $\mathcal O(n^2)$, we can update its value in $\mathcal O(n)$. In Section \ref{sec:conc}, we discuss how this could be improved even further using an edge list instead of adjacency matrix. Additionally, this type of improvement could also be used to efficiently calculate alternative CP objective functions, i.e., Eq. (4) in \cite{BORGATTI2000} where the number of core-periphery edges are excluded and only core-core and periphery-periphery edges are used.

\subsection{Time complexity}
We precisely characterize the complexity of one pass of the modified algorithm. Since there are $n$ nodes in the network, initializing the CP labels is $\mathcal O(n)$, as is randomly ordering the nodes. Computing $T(\bA,\bc)$ is $\mathcal O(n^2)$, but $T(\bA,\bc')$ can be updated with only $\mathcal O(n)$ operations. This must be computed for each node such that one pass of the for loop takes $\mathcal O(n^2)$ operations. Note that if we re-computed $T(\bA,\bc')$ from scratch at every iteration, then the pass would require $\mathcal O(n^3)$ operations. Thus, our improvement yields an order-of-magnitude computational speed-up.

\subsection{Convergence guarantees}

In this sub-section, we discuss some of the theoretical properties of the proposed algorithm. Algorithm 1 is a greedy algorithm in the sense that it makes the locally optimal decision. Ideally, we hope that this algorithm adequately approximates the global optimum in \eqref{eq:opt}.

We first establish monotone ascent to a local maximum for Algorithm 1. To do so, we first must define what ``local'' means. Let $N(\bc)$ be a neighborhood of label $\bc$ defined as
$$
    N(\bc)
    =\left\{\bc':\sum_{i=1}^n\mathbb I(c_i'\neq c_i)=1\right\}.
$$
where $\mathbb I(\cdot)$ is the indicator function.
In words, this is the set of all CP labels that differ from $\bc$ by only one node assignment. In the following proposition, we show that Algorithm 1 converges to a local optimum by this definition.\\

\noindent 
{\sc Proposition 1.} {\it Let $\hat\bc_1$ be the labels returned by Algorithm 1. Then for all $\bc\in N(\hat\bc_1)$, $T(\bA,\hat\bc_1)\geq T(\bA,\bc)$, i.e., Algorithm 1 enjoys monotone ascent to a local maximum.}\\ 

\noindent
A brief proof is given in Appendix A. We emphasize that since Algorithm 1 updates the labels only when $T(\bA,\bc')$ is {\it strictly greater} than $T(\bA,\bc)$, cycling among equal-valued CP labels cannot occur. Proposition 1 ensures that the labels returned by Algorithm 1 yield a local maximum of $T(\bA,\bc)$. We would like to have some guarantee, however, of how close this result is to the {\it global} maximum. In many proofs that an algorithm achieves the global maximum, the {\it sub-modularity} of the objective function is required, e.g., \cite{kempe2003maximizing}. Unfortunately, the BE metric is not sub-modular, meaning that there are no greedy optimality guarantees, and we can only prove the monotone ascent property of Proposition 1.

\begin{figure}
    \centering
    \includegraphics[width=0.9\linewidth]{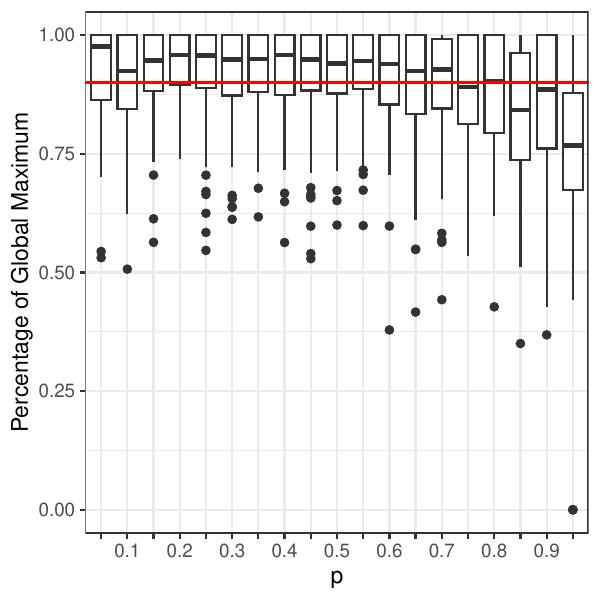}
    \caption{Ratio of the maximum value of the Borgatti and Everett metric returned by Algorithm 1 with the true global optimum for Erdos-Renyi networks with different values of $p$. The red horizontal line denotes 90\%.}
    \label{fig:brute}
\end{figure}

For small $n$, however, we can empirically compare Algorithm 1 to the global optimum by using a brute force approach. Specifically, we let $n=20$ and generate Erdos-Renyi (ER) networks with $P_{ij}=p$ for all $i,j$ \citep{Erdos1959}. We use ER networks because they do not have an intrinsic CP structure, making the optimization task more difficult. Since $n=20$, there are $2^{20}\approx 10^6$ possible CP labels and we evaluate each one (exhaustive enumeration) to find $\hat\bc=\arg\max_{\bc}\{T(\bA,\bc)\}$, i.e., the global optimum. We vary $p=0.05,0.10,\dots,0.95$ and find $T(\bA,\hat\bc_1)/T(\bA,\hat\bc)$, i.e., the ratio of the optimum value returned by Algorithm 1 to the true global optimum. In Figure \ref{fig:brute}, we report the box-plot of these ratios over 100 Monte Carlo (MC) replicates. In general, the labels returned by Algorithm 1 exceed 90\% of the global optimum. As the network density increases, however, the performance slightly decreases. This result is not too concerning as most real-world networks are relatively sparse, and in these regimes (small $p$), the proposed algorithm performs best. Thus, these results show that the proposed method yields CP labels close to the global optimum. We stress that these results apply to Algorithm 1 in general, and are not related to the \texttt{greedyFast} computational improvements. Finally, we estimate that for $n=30$, a brute-force approximation (without parallelization) would take approximately ten hours to complete, meaning that this approach becomes infeasible for networks much larger than this.

\section{Simulation study}\label{sec:sim}
In this section, we compare the proposed \texttt{greedyFast} with the naive implementation (\texttt{greedyNaive}) as well as an existing algorithm to identify CP structure on synthetic networks. Specifically, \texttt{cpnet} is a popular Python package for various network tasks related to CP structure and can be used in R with the \texttt{reticulate} package \citep{reticulate}. This implementation also uses a Kernighan-Lin-type algorithm to optimize the BE metric. Please see the Appendix for more details. We chose this algorithm as the main competitor because it is freely available online and can easily be implemented on Mac. On the other hand, UCINET \citep{everett2002ucinet}, for example, is designed for Windows, making use on a Mac more difficult. Additionally, we do not compare with other CP algorithms because the specific goal of this work is proposing a new algorithm for optimizing the BE metric.

We generate networks with a known CP structure to compare the labels returned by the algorithms with the ground-truth. In the following sub-sections, we describe the data-generating models, but for now, let $\bc^*$ be the true labels and $\hat\bc$ be the estimated labels returned by one of the algorithms. Then the classification accuracy can be defined as 
$$
    \frac1n\sum_{i=1}^n \mathbb I(\hat c_i= c_i^*).
$$
We report the classification accuracy and run-time for each algorithm applied a single time to each network. All experiments are carried out on a 2024 Mac Mini with 16 GB of memory, and the results are averaged over 100 MC samples. The R code to implement \texttt{fastGreedy}, as well as replicate the following simulation study, is available on the author's GitHub: \url{https://github.com/eyanchenko/fastCP}.

\subsection{Stochastic Block Model CP structure}
First, we generate networks according to a 2-block stochastic block model (SBM) \citep{holland1983stochastic}. We let $P_{ij}=p_{c^*_i,c^*_j}$ for $i,j\in\{1,2\}$ with $p_{11}>p_{12}=p_{21}>p_{22}$ to ensure the networks have a CP structure where we use a slight abuse of notation as now $c_i^*=2$ means that node $i$ is in the periphery. Unless otherwise noted, we fix $p_{11}=2p_{12}$, i.e., the core-core edge probability is twice the core-periphery edge probability, and $k=0.1n$, i.e., 10\% of the nodes are in the core.

\begin{figure}
     \centering
     \begin{subfigure}[b]{0.9\textwidth}
         \centering
         \includegraphics[width=\textwidth]{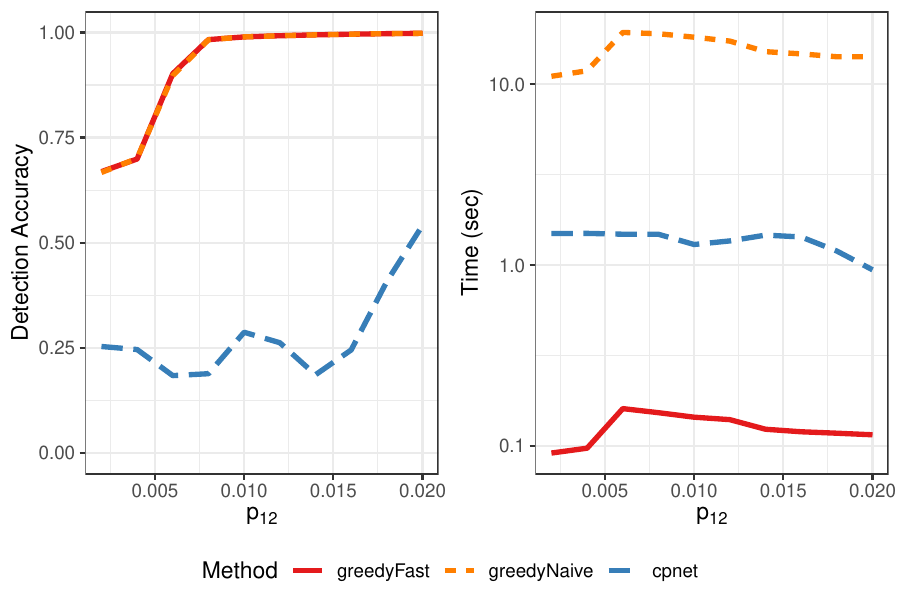}
         \caption{Accuracy and computing time for varying $p_{12}$.}
         \label{fig:SBM_p12}
     \end{subfigure}
     \\
     \begin{subfigure}[b]{0.9\textwidth}
         \centering
         \includegraphics[width=\textwidth]{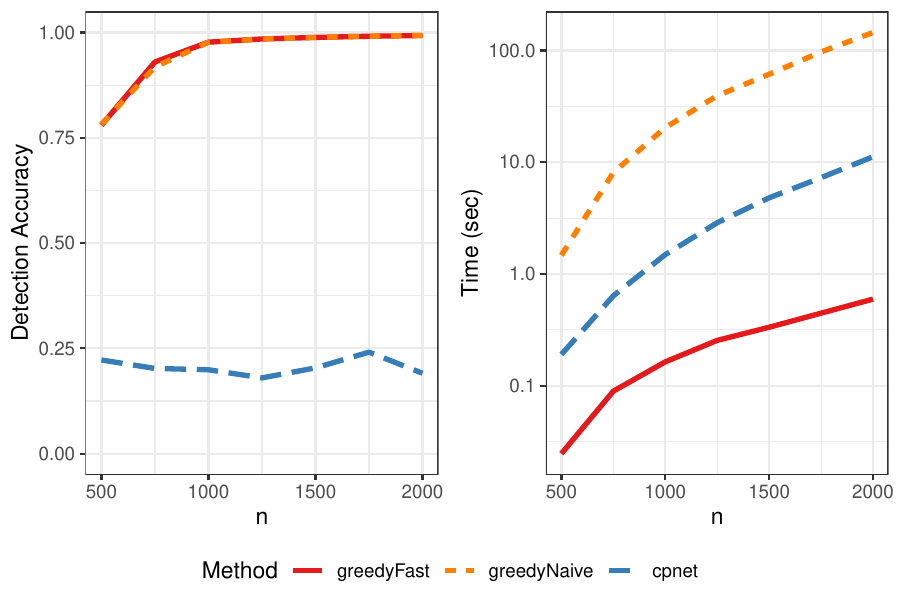}
         \caption{Accuracy and computing time for varying $n$.}
         \label{fig:SBM_n}
     \end{subfigure}
        \caption{Core-periphery identification results on SBM networks.}
        \label{fig:SBM}
\end{figure}

In the first set of simulations, we fix $n=1000$, $p_{22}=n^{-1}=0.001$ and vary $p_{12}=0.002$, $0.004$, $\dots$, $0.02$. As $p_{12}$ increases, so too does the strength of the CP structure in the network. The results are in Figure \ref{fig:SBM_p12}. \texttt{greedyFast} has a much larger detection accuracy and is approximately one order of magnitude faster than \texttt{cpnet} for all values of $p_{12}$. As expected, \texttt{greedyNaive} yields identical accuracy results, but the proposed algorithmic improvement significantly decreases computing time. In the Appendix, we also re-run \texttt{greedyFast} on the same network multiple times to understand the effect of the stochasticity in the algorithm. In general, we find very small variance in terms of accuracy and run-time.

Next, we fix $p_{12}=0.005$ and $p_{22}=0.001$ and vary $n=500,750,\dots,2000$ with results in Figure \ref{fig:SBM_n}. Again, \texttt{greedyFast} clearly outperforms \texttt{cpnet} with superior accuracy, and is much faster than both the naive implementation and \texttt{cpnet} algorithm. It also appears that \texttt{greedyFast} and \texttt{greedyNaive} are more stable in both settings as the classification accuracy monotonically increases with $p_{12}$ and $n$, unlike that of \texttt{cpnet}. Finally, in the Appendix, we present a log-log plot of the runtime with $n$ to verify the algorithm's computational complexity.

\subsection{Degree-corrected Block Model CP structure}
\begin{figure}
     \centering
     \begin{subfigure}[b]{0.9\textwidth}
         \centering
         \includegraphics[width=\textwidth]{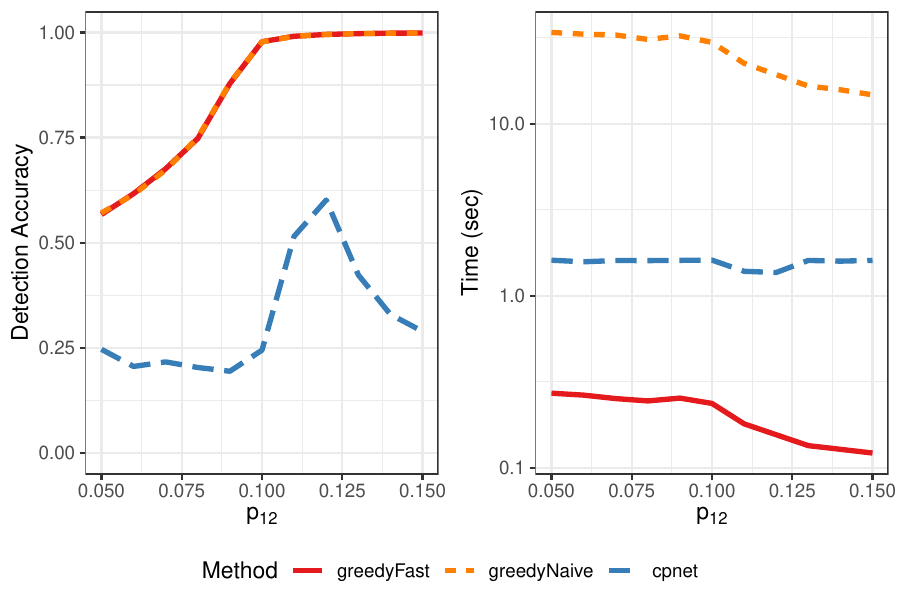}
         \caption{Accuracy and computing time for varying $p_{12}$.}
         \label{fig:DCBM_p12}
     \end{subfigure}
     \\
     \begin{subfigure}[b]{0.9\textwidth}
         \centering
         \includegraphics[width=\textwidth]{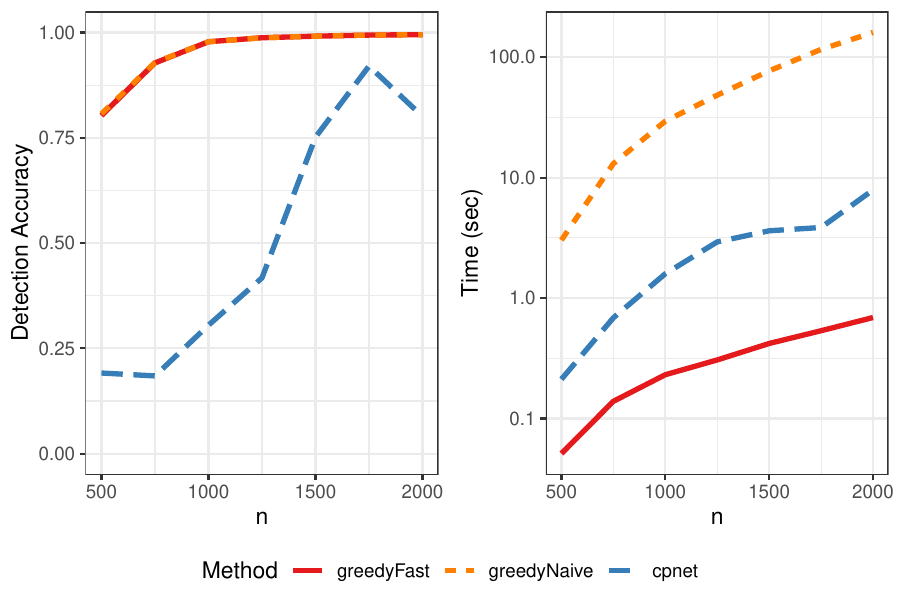}
         \caption{Accuracy and computing time for varying $n$.}
         \label{fig:DCBM_n}
     \end{subfigure}
        \caption{Core-periphery identification results on DCBM networks.}
        \label{fig:DCBM}
\end{figure}

In the second set of simulations, we generate networks from a degree-corrected block model (DCBM) \citep{karrer2011stochastic}. Namely, we introduce weight parameters $\btheta=(\theta_1,\dots,\theta_n)^\top$ to allow for degree heterogeneity among nodes, and generate networks with $P_{ij}=\theta_i\theta_jp_{c_i^*,c_j^*}$. If $p_{11}>p_{12}>p_{22}$, then we expect the resulting networks to exhibit a CP structure. We sample $\theta_i\stackrel{\text{iid.}}{\sim}\mathsf{Uniform}(0.6,0.8)$ and fix $p_{22}=0.05$. First, we fix $n=1000$ and vary $p_{12}=0.05, 0.06,\dots, 0.15$, with results in Figure \ref{fig:DCBM_p12}. Second, we fix $p_{12}=0.10$ and vary $n=500,750,\dots,2000$. As in the SBM examples, \texttt{greedyFast} yields the same accuracy as \texttt{greedyNaive}, but is markedly faster. It is also significantly faster and more accurate than the \texttt{cpnet} implementation.

\section{Real-data analysis}\label{sec:real}

Lastly, we apply \texttt{greedyFast} to thirteen real-world networks. We consider data sets of a variety of domains, sizes and densities to ensure the broad applicability of the method, including: {\it UK Faculty} (number of nodes $n=81$, average edge probability $\bar A=0.18$) \citep{nepusz2008fuzzy}, {\it Email} ($n=167, \bar A=0.23$) \citep{michalski2011matching}, {\it British MP} ($n=381, \bar A=0.08$) \citep{greene2013producing}, {\it Congress} ($n=475, \bar A=0.09$) \citep{fink2023centrality}, {\it Political Blogs} ($n=1224, \bar A=0.02$) \citep{adamic::2005aa}, {\it DBLP} ($n=2203, \bar A=0.47$) \citep{gao2009graph, ji2010graph}, {\it Hospital} ($n=75, \bar A=0.41$) \citep{Vanhems:2013}, {\it School} ($n=242, \bar A=0.29$) \citep{stehle2011high}, {\it Bluetooth} ($n=672, \bar A=0.10$) \citep{sapiezynski2019interaction}, {\it Biological 1, 2, 3} ($n=2220,3289,4040, \bar A=0.02, 0.02, 0.01$, respectively) \citep{cho2014wormnet}, and {\it Facebook} ($n=4039, \bar A=0.10$) \citep{leskovec2012learning}. All data was retrieved from \citep{snapnets, nr, csardi2013package}. For each network, we remove all edge weights, directions, time stamps, self-loops, etc. We also only consider two groups of researchers in the DBLP network as in \cite{senguptapabm}. 
We run \texttt{greedyFast} and the \texttt{cpnet} algorithm to identify the optimal CP labels, and record $T(\bA,\hat\bc)$, the value of the BE metric evaluated at the optimal labels returned by each algorithm, the number of core nodes in these optimal labels and the computing time. Since ground-truth labels are unknown in these networks, achieving a higher objective value $T(\bA,\hat\bc)$ is the main criterion to judge which algorithm returns superior labels. We run each algorithm 10 times and keep the results corresponding to the largest value of the objective function (along with the total run time). Results are reported in Table \ref{tab:real}. In this section, we run the \texttt{cpnet} algorithm directly in Python to eliminate any time that might be added from using the \texttt{reticulate} package. In general, however, we found virtually no difference in the run-time or performance when using the function in R or Python.

Overall, the results demonstrate strong performance of the proposed method. In about half of the networks, the optimal value of the objective function is similar between the two algorithms, but \texttt{greedyFast} is consistently one to two orders of magnitude faster. On the remaining networks ({\it British MP}, {\it Political Blogs}, {\it DBLP}, {\it Biological 1 - 3}, {\it Facebook}), the proposed algorithm is not only faster than that of \texttt{cpnet} but also yields noticeably superior labels in terms of the value of $T(\bA,\hat\bc)$. Moreover, on {\it DBLP}, \texttt{greedyFast} is $217/0.630\approx 344$ times faster than the competing method. Additionally, our method appears to converge to a local optimum for all networks, but the \texttt{cpnet} algorithm seems not to in {\it British MP}, {\it Biological 2} and {\it Facebook}. Even after re-running the algorithm on these networks, the result did not improve. This could be a result of the \texttt{cpnet} algorithmic design where node labels are fixed after they have been swapped. In summary, the proposed algorithm yields better approximations for the global maximum at a fraction of the computation time.

\begin{table}[]
    \centering
    \begin{tabular}{lc|ccc}
        Network & Algorithm & $T(\bA,\hat\bc)$ & $k$ & Time \\\hline 
        UK Faculty  & \texttt{greedyFast} & 0.26 & 16 & 0.001\\
        & \texttt{cpnet} & 0.25 & 20 & 0.364\\\hline
        
        Email  & \texttt{greedyFast} & 0.42 & 22 & 0.004\\
        & \texttt{cpnet} & 0.42 & 14& 0.481\\\hline
        
        British MP & \texttt{greedyFast} & 0.25 & 61 & 0.021\\
        & \texttt{cpnet} & 0.04 & 1& 0.970\\\hline
        
        Congress  & \texttt{greedyFast} & 0.18 & 67 & 0.044\\
        & \texttt{cpnet} & 0.18 & 66 & 0.092\\\hline
        
        Political Blogs  & \texttt{greedyFast} & 0.21 & 91 & 0.249\\
        & \texttt{cpnet} & 0.14 & 323 & 28.8\\\hline
        
        DBLP & \texttt{greedyFast} & 0.26 & 529 & 0.630\\
        & \texttt{cpnet} & 0.20 & 856 & 217 \\\hline
        
        Hospital & \texttt{greedyFast} & 0.44 & 22 & 0.001\\
        & \texttt{cpnet} & 0.44 & 24 & 0.017 \\\hline
        
        School & \texttt{greedyFast} & 0.22 & 52 & 0.009\\
        & \texttt{cpnet} & 0.22 & 53 & 0.075 \\\hline
        
        Bluetooth & \texttt{greedyFast} & 0.17 & 127 & 0.063\\
        & \texttt{cpnet} & 0.17 & 128 & 3.17 \\\hline
        
        Biological 1 & \texttt{greedyFast} & 0.13 & 402 & 1.152\\
        & \texttt{cpnet} & 0.03 & 14 & 171 \\\hline
        
        Biological 2 & \texttt{greedyFast} & 0.13 & 269 & 3.71\\
        & \texttt{cpnet} & 0.00 & 3287 & 546 \\\hline
        
        Biological 3 & \texttt{greedyFast} & 0.10 & 376 & 6.44\\
        & \texttt{cpnet} & 0.07 & 1472 & 1007 \\\hline
        
        Facebook & \texttt{greedyFast} & 0.10 & 291 & 8.121\\
        & \texttt{cpnet} & 0.01 & 2 & 1036 \\\hline
    \end{tabular}
    \caption{Results of real-data analysis using \texttt{greedyFast} and \texttt{cpnet}. $T(\bA,\hat\bc)$: value of objective function at optimal labels returned by the algorithm; $k$: number of nodes assigned to the core in the optimal labels; Time: computing time (seconds).}
    \label{tab:real}
\end{table}

\section{Conclusion}\label{sec:conc}

In this work, we detail a greedy, label-switching algorithm to quickly identify core-periphery (CP) structure in networks. In comparisons with an existing algorithm, our greedy approach demonstrates superior core-periphery label identification, while also being theoretically guaranteed to terminate at a local maximum. Additionally, by noting the mathematical relationship between the proposed and current CP labels, our implementation greatly reduces the computation time. Indeed, on both synthetic and real-world networks, we typically observe an order of magnitude improvement in algorithmic run-time.

There are several interesting avenues for future work. Precise theoretical results surrounding the conditions needed to obtain a global optimum would be worthwhile. Additionally, we focused on algorithmic speed-ups stemming from mathematical properties of the objective function. Considering more efficient network storage and data structures, however, could also improve the efficiency of the algorithm. For example, we currently compute $\sum_{k=1}^i A_{ik}\Delta_{ik}$ in $\mathcal O(n)$ operations. But on sparse networks where $\mathsf{E}(A_{ij})=\rho_n P_{ij}$ for some sparsity parameter $\rho_n\to0$, if we use an edge list, then the sum could be computed in $\mathcal O(n\rho_n)$ operations, leading to potentially large gains. Lastly, our algorithmic approach could be applied to generalizations of the BE metric, e.g., \cite{estevez2025revising}.

\paragraph{Acknowledgments}
We would like to thank Sadamori Kojaku for help understanding the \texttt{cpnet} algorithm. We also greatly appreciate the feedback and comments from the reviewers and editors.

\paragraph{Funding Statement}
None to report.

\paragraph{Competing Interests}
None to report.

\paragraph{Data Availability Statement}
All data used in this study is freely available online.

\paragraph{Ethical Standards}
N/A

\paragraph{Author Contributions}
Conceptualization: EY, SS. Methodology: EY. Coding: EY. Writing original draft: EY. Mentoring: SS. All authors approved the final submitted draft.

%\endnote in some journals will behave like \footnote; and \printendnotes will not output anything. 

%\printbibliography[prenote={preamble}]
\printbibliography
\appendix

\section*{Appendix}
\subsection*{Proof of Proposition 1.}
{\it Let $\hat\bc_1$ be the labels returned by Algorithm 1. Then for all $\bc\in N(\hat\bc_1)$, $T(\bA,\hat\bc_1)\geq T(\bA,\bc)$, i.e., Algorithm 1 converges to a local optimum.}\\

\noindent
{\it Proof.} Recall the steps of Algorithm 1. Then the result is immediate. Let $\hat\bc_1$ be the labels that Algorithm 1 converged to. Assume, for contradiction, that there is some $\bc'\in N(\hat\bc_1)$ such that $T(\bA,\bc') > T(\bA,\hat\bc_1)$. But this is a contradiction because if $T(\bA,\bc') > T(\bA,\hat\bc_1)$, then Algorithm 1 would have swapped the differing labels such that $\hat\bc_1=\bc'$. Therefore, it must be that $T(\bA,\hat\bc_1) \geq T(\bA,\bc')$ for any $\bc'\in N(\hat\bc_1)$. $\square$

\subsection*{cpnet implementation}
We now provide more details on the \texttt{cpnet} algorithm to find the optimal CP labels. This algorithm uses a Kernighan-Lin \citep{Kernighan1970} approach. Specifically, the CP labels, $\bc$, are first randomly initialized to the core or periphery with equal probability. Then, for each node, the label of the node is switched and the change in the objective function ($T(\bA,\bc)$) is calculated. After swapping all $n$ nodes, the switch which led to the largest increase in the objective function value is kept. Then this node's label is held fixed for the remainder of the algorithm and the process repeats. The algorithm terminates when the improvement in the objective function is less than or equal to $\varepsilon=10^{-7}$, or if all nodes have been checked.

There are some key similarities and differences between this algorithm and \texttt{greedyFast}. First, both methods use a local-greedy search by swapping the label of a single node. On the other hand, \texttt{greedyFast} does not fix a node's label once it has been swapped, and the node order is randomized each loop through the algorithm; in \texttt{cpnet}, this order is fixed throughout. Additionally, \texttt{greedyFast} swaps a node label immediately if it increases the objective function, not waiting until all nodes have been checked. It also has a different stopping condition, terminating the algorithm when all $n$ nodes have been swapped and none of them led to an increase in the objective function. Additionally, the default implementation in \texttt{cpnet} re-runs the algorithm ten times with different initial labels, and then keeps only the result corresponding to the largest objective function value. This was changed to a single run for the simulation studies.

\subsection*{Sensitivity analysis}
\begin{figure}
    \centering
    \includegraphics[width=0.95\linewidth]{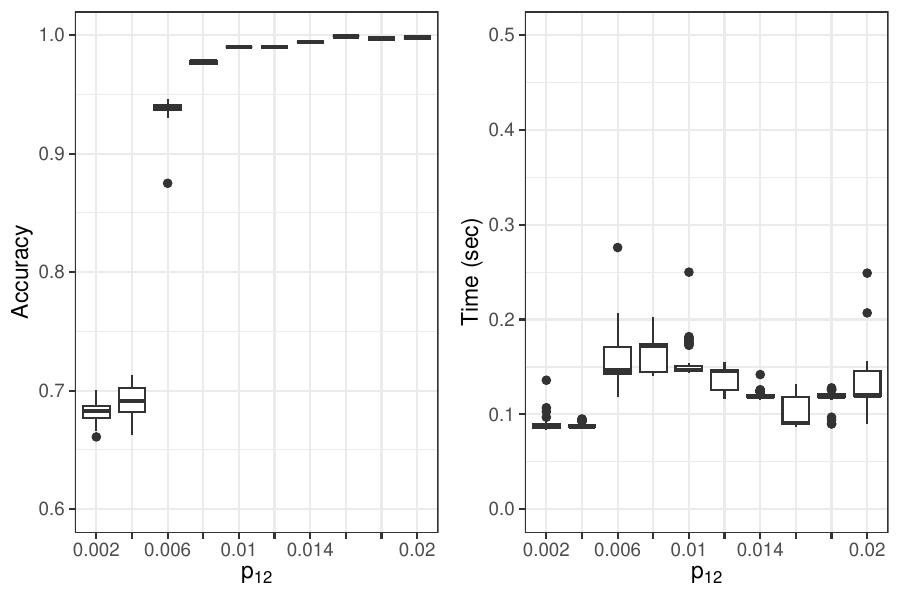}
    \caption{Accuracy and computing boxplots for sensitivity analysis.}
    \label{fig:app}
\end{figure}

We briefly study the effect of the node order randomization in Algorithm 1.
In the implementation of both \texttt{greedyFast} and \texttt{greedyNaive}, we randomly initialize the CP labels, $\bc$, as well as randomly shuffle the node order when swapping the CP membership assignments. To understand the effect of this stochasticity, we generate networks as in Section 4.1, i.e., SBM with $n=1000$, $p_{11}=2p_{12}$, $p_{22}=0.001$, $k=0.1n$, and vary $p_{12}=0.002, 0.004,\dots,0.02$. Instead of generating 100 networks for each parameter combination, we instead generate a single network but apply \texttt{greedyFast} to this same network 100 times. We report the accuracy and run-time in box-plots in Figure \ref{fig:app}. Clearly, there is very little variance in the results, indicating minimal sensitivity to the randomization in Algorithm 1.

\subsection*{Runtime complexity}
\begin{figure}
    \centering
    \includegraphics[width=0.95\linewidth]{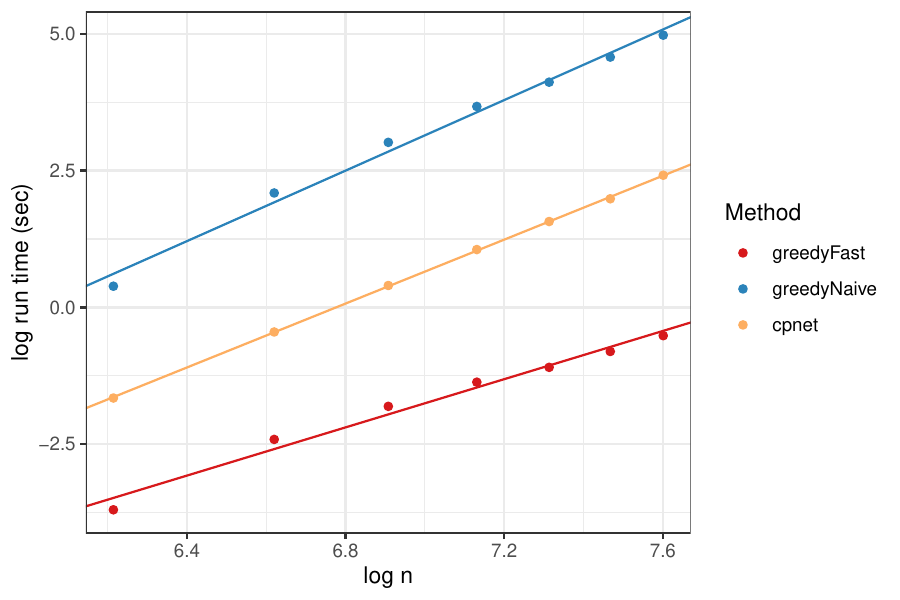}
    \caption{Log runtime against log $n$ for the SBM simulations.}
    \label{fig:app_log}
\end{figure}

We further compare the runtime of the various algorithms. In Figure \ref{fig:app_log}, we plot the log runtime against the log number of nodes for the SBM synthetic network simulations. We also regress log runtime on log $n$ for each algorithm and find that the slope is 3.2, 2.9 and 2.2 for \texttt{greedyNaive}, \texttt{cpnet} and \texttt{greedyFast}, respectively. For the DCBM simulations, the respective fitted slopes are 2.8, 2.5 and 1.8. These results further validate the computational improvements of the proposed implementation.

\end{document}